\documentclass[twocolappendix,iop,apj]{emulateapj}
\usepackage{amsfonts,amsmath,graphicx,natbib,url,hyperref,nicefrac}
\hypersetup{colorlinks=true, urlcolor=blue, citecolor=blue}

\usepackage{amsmath}
\usepackage{tabularx}
\usepackage{threeparttable}
\usepackage[usenames,dvipsnames]{color}
\usepackage{comment}

\shorttitle{Afterglow Plateaus from Stratified Ejecta} 
\shortauthors{Duffell et al}

\begin{document}

\title{High-Resolution Numerical Calculations of GRB Afterglow Plateaus arising from Stratified Ejecta}

\def\pur{1}
\def\cal{2}

\author{Paul C. Duffell, Ranadeep Ghosh Dastidar}
\affiliation{Purdue University, 525 Northwestern Avenue, West Lafayette, IN 47907}
\email{pduffell@purdue.edu}

\begin{abstract}

Since the discovery of plateaus in GRB afterglows by Swift, they have been modeled predominantly by late-time energy injection.  However, many studies have suggested that the plateau may be modeled by an early phase before reverse shock crossing (either coasting with constant Lorentz factor or decelerating very slowly as the reverse shock crosses the ejecta).  The slope of the early plateau provides some constraints the stratification of the fastest-moving ejecta, which could be the ejecta responsible for the prompt emission.  However, numerical studies typically do not model the jet as a stratified outflow; the reason being the extremely high resolution required in order to accurately evolve this tiny amount of highly relativistic material.  In this study, we perform high-resolution numerical calculations ($\Delta r/r \lesssim 10^{-5}$) verifying that a stratified ejecta structure can indeed produce an afterglow plateau in both wind ($k=2$) and ISM ($k=0$) environments, and computing break times explicitly.  We evolve the relativistic hydrodynamics using the \texttt{JET} code, and post-process this to compute the afterglow using the \texttt{Firefly} code.  Our results show that a stratified ejecta structure (which should generically be present in GRB jets) is sufficient to explain GRB afterglows, and the plateau slopes can be used to constrain the ejecta stratification.  We additionally provide precise measured scalings for the plateau duration as a function of the characteristic Lorentz factor of the ejecta.  The long duration of typical plateaus requires a very modest characteristic Lorentz factor for the ejecta ($\gamma_0 \sim 10-50$), in agreement with other afterglow plateau models.

\end{abstract}


\section{Introduction} \label{sec:intro}

Since their first detection by Swift, Plateaus in GRB afterglows have been a long-running mystery in the GRB community \citep{2006ApJ...642..389N, 2006ApJ...642..354Z}.  They are typically detected in X-Rays (sometimes optical if follow-up is available) and typically take place starting from the end of the prompt emission, lasting at least for minutes but sometimes up to a day ($10^3 - 10^5$ seconds) after which there is a break in the light curve, usually transitioning to a sufficiently steep decline to model using the traditional Blandford-McKee blastwave model \citep{1976PhFl...19.1130B}

The mystery is not just what causes the plateau itself; the big question is why we do not see brighter emission from a relativistic ($\gamma \sim 100$) blastwave at these early times.  There is no obvious reason the Blandford-McKee phase should not extend back in time to periods when the Lorentz factor of the shock is much higher.  Using standard afterglow modeling and typical jet parameters, the Lorentz factor of the jet at the end of the plateau is around $\gamma \sim$ a few tens, depending on the plateau duration \citep{2007MNRAS.379..331P}.  So why do we not see this evolution at earlier times, given that we expect these jets to have a much higher initial Lorentz factor (at least $\gamma \sim 100$ is typically necessary to blueshift their emission into the gamma rays we see without producing an opaque pair plasma \citep{2001ApJ...555..540L, 2019MNRAS.486.1563M})?

A number of models have been suggested over the past few decades, none of which are perfect solutions to this problem.  The most commonly accepted explanation is late-time energy injection \citep{2006ApJ...642..354Z,
2006MNRAS.366.1357P}.  This model can certainly reproduce the correct light curves, but that is of course by construction; if we are free to decide how energy is injected over time, a sufficiently creative theorist can produce almost any light curve they like.  

An important problem with the late-time injection model is that it is hard to imagine what sort of GRB engine lasts for $10^3 - 10^5$ seconds \citep{2006ApJ...642..354Z}, especially when the GRB itself may only last for 10 seconds, which is thought to be associated with the accretion timescale of the gas powering the jet.  This requires invoking longer-duration mechanisms such as fallback accretion or magnetars \citep{2008Sci...321..376K, 2011MNRAS.413.2031M}.  As most of the energy from the jet would need to be injected at late times, this requires the afterglow jet is essentially generated by a different mechanism than the less energetic $\gamma \gtrsim 100$ jet that produced the original gamma rays.

Because of these challenges, over the past several decades there have been many other models suggested as alternatives to late-time energy injection.  For example, a plateau could naturally be generated by geometric effects if the jet is not viewed perfectly on-axis \citep{2020MNRAS.492.2847B}.  Another example that has arisen a number of times in the literature associates the plateau with the early ``coasting" phase of the jet, before it has swept up enough mass to decelerate significantly.  To our knowledge this was first suggested by \cite{2012ApJ...744...36S}.  Of course, associating the plateau with the coasting phase demands that the jet Lorentz factor was $10-50$ at all times before the plateau, which is again in tension with the evidence for $\gamma \gtrsim 100$ outflows coming from the prompt emission.  This difficulty was addressed by \cite{2015ApJ...806..205D}, who envisioned the afterglow jet as a by-product of a collision between a highly relativistic $\gamma \gtrsim 100$ outflow and a modestly relativistic $\gamma \sim 10$ outflow, citing evidence from numerical calculations of jets propagating out of collapsars \citep[e.g.][]{2009ApJ...698.1261Z}.  The coasting model was also independently suggested by \cite{2022NatCo..13.5611D}.

Unfortunately, coasting-jet models have a major drawback, which is that they require a wind-like external density profile $\rho = A/r^2$ in order to guarantee a shallow plateau.  For a simple ISM-like profile ($\rho = $ constant), the model predicts a rising flux $F \propto t^2$, which is almost never seen in the plateau phase, except of course for the case of off-axis GRBs such as GRB 170817A \citep{2017Sci...358.1579H, 2017ApJ...848L..20M}).  It is expected that at least some portion of GRBs should have some ISM-like surrounding gas, especially at sufficiently large radii probed by later plateau breaks (the wind should eventually terminate at some radius, transitioning to ISM).  Power-law fits of GRB afterglows find examples that are consistent with $k=0$ as well as $k=2$ \citep{2011A&A...526A..23S}.  Even if a small fraction of GRB afterglows exhibit a uniform-density circumburst environment, this is likely sufficient to rule out the coasting model, as we almost never see a $t^2$ rising phase in GRB afterglows.

It should be noted that the coasting model is actually a limiting case of a class of models that decelerate according to a shallow power-law as the reverse shock propagates through the ejecta.  In other words, ``pure coasting" with a constant Lorentz factor assumes a piston-like solution where the ejecta behaves like a constant-velocity boundary condition, which one achieves in the limit that the ejecta density gradient is infinitely steep.  If we relax that assumption, we find a family of solutions where a reverse shock propagates through a power-law ejecta structure.  This is essentially the ultra-relativistic version of the Chevalier self-similar solutions used extensively to characterize the fastest-moving ejecta in supernovae \citep{1982ApJ...258..790C}.  The scalings for these ultra-relativistic solutions were first derived by \cite{2000ApJ...535L..33S}, and they have been used to model GRB plateaus in the past, without any additional energy injection \cite{2006MNRAS.366L..13G}.  More detailed calculations of these scalings were provided by \cite{2026MNRAS.549ag833S}, who argued that these self-similar solutions could be used as a model to explain the GRB afterglow plateau, providing a shallow decay in the X-rays for any external density profile (in particular for both $k=0$ and $k=2$ cases).

The only assumption in this case is that the density structure is ``stratified", that is not all the mass is moving at a single Lorentz factor, but there is a power-law tail of material moving at much larger Lorentz factors.  But this is not really much of an assumption; outflows should generically be stratified.  In fact, it would be somewhat pathological to assume all of the ejecta moves at exactly the same Lorentz factor.  Thus, stratified outflows are an expected outcome of any jet launching mechanism, and the scalings of \cite{2000ApJ...535L..33S} show that they would naturally generate shallow plateaus in GRB afterglows.  A number of GRB afterglows have been modeled with this specific interpretation in mind\footnote{As a semantic note, stratified-ejecta or coasting phase models can be (and often are) considered a form of ``late-time energy injection", as the unshocked ejecta injects energy into the shocked region as time progresses.  In this study, we make a semantic separation between this process and models that require late-time injection directly by an engine, such as a magnetar.  The reason being that stratified-ejecta models typically assume the same outflow that powers the GRB itself is also responsible for the afterglow, potentially enabling predictions for correlations between the prompt and afterglow phases. } \citep[e.g.][]{2015ApJ...814....1L, 2018ApJ...862...94L, 2025ApJ...987..178S}.

Much like in the coasting model, the end of the plateau is associated with the time the reverse shock crosses the ejecta, requiring that the typical Lorentz factor of the bulk of the ejecta be in the low-to-mid tens, $\gamma_0 \sim 10-30$, but the stratified outflow naturally assumes some of the material still has much higher Lorentz factor, potentially responsible for the prompt emission itself.  However, \cite{2026MNRAS.549ag833S} argued that the reverse shock crossing can be delayed sufficiently to increase the inferred Lorentz factor at the break to around $\sim 100$ (in other words, most of the quoted Lorentz factors here are estimates, whereas \cite{2026MNRAS.549ag833S} argued for a specific coefficient in front).

In this study, we confirm the above analytical predictions with the first-ever numerical calculations, both in the relativistic hydrodynamics and the calculation of the synchrotron radiation from the shock.  We compute synthetic light curves from our hydrodynamical output and show that shallow, decaying X-ray plateaus are indeed robustly generated, and empirically measure the scaling law connecting the Lorentz factor of the ejecta to the plateau duration.  We find for typical GRB parameters, observed plateau breaks imply typical bulk Lorentz factors of $\sim 10-50$, in agreement with previous analytical estimates.  However, the power-law tail of stratified material guarantees some fraction of the jet always has $\gamma > 100$.

\section{Analytical Considerations}
\label{sec:anly}
Analyical predictions for forward shock evolution in a stratified flow were first computed by \cite{2000ApJ...535L..33S}, and have also been derived in detail by \cite{2026MNRAS.549ag833S}.  We repeat their main conclusions here.

A stratified density profile can be characterized by the scaling of ``external mass" with Lorentz factor, i.e. how much mass is moving with Lorentz factor in excess of $\gamma$:

\begin{equation}
    M_{> \gamma} = M (\gamma/\gamma_0)^{-s}
\end{equation}

Alternatively this can be described in terms of the energy stratfication:

\begin{equation}
    E_{> \gamma} \propto (\gamma/\gamma_0)^{1-s}
\end{equation}

This stratified ejecta collides with an external medium with density profile of

\begin{equation}
    \rho = A/r^k.
\end{equation}

As the two fluids collide, a forward shock (FS) and reverse shock (RS) form.  So long as the RS has not crossed the ejecta, the FS Lorentz factor scaling with time can be estimated as

\begin{equation}
    \gamma^2 \sim E/M_{\rm swept} \sim \frac{ E_{> \gamma} }{ \int r^2 \rho(r) dr |_{r=ct} } \sim \frac{\gamma^{1-s}}{t^{3-k}}.
\end{equation}

Thus, the forward shock's Lorentz factor scales with time as:

\begin{equation}
    \gamma(t) \propto t^{-\frac{3-k}{1+s}}
    \label{eqn:anlygam_lab}
\end{equation}

The above equation is described in the ``lab frame".  In observer time, $t_{\rm obs} \sim t/\gamma^2$, this is

\begin{equation}
    \gamma(t_{\rm obs}) \propto t_{\rm obs}^{-\frac{3-k}{7+s-2k}}.
    \label{eqn:anlygam}
\end{equation}

Choosing $s=1$ effectively enforces negligible mass, recovering the Blandford-McKee \citep[BMK;][]{1976PhFl...19.1130B} solution:

\begin{equation}
    \gamma(t) \sim (t/t_{\rm Sedov})^{-\frac{3-k}{2}}
\end{equation}

\begin{equation}
    \gamma(t_{\rm obs}) \sim (t_{\rm obs}/t_{\rm Sedov})^{-\frac{3-k}{8-2k}}
\end{equation}

where $t_{\rm Sedov}$ is determined by the time when the flow becomes trans-relativistic $\gamma \sim 1$, when the swept-up rest mass is comparable to the isotropic equivalent energy of the jet:

\begin{eqnarray}
    t_{\rm Sedov} \equiv (E_{\rm iso}/A)^{\frac{1}{3-k}}.
\end{eqnarray}

The transition from plateau to Blandford-McKee should occur when the forward shock decelerates to $\gamma \sim \gamma_0$, at observer time

\begin{equation}
    t_{\rm p} \sim t_{\rm Sedov} \gamma_0^{-\frac{8-2k}{3-k}}
\end{equation}

For a uniform density ISM, this scaling looks like

\begin{equation}
    t_{\rm p} \sim 10^4 s \left( \frac{E_{\rm iso}}{10^{53} \rm ergs} \right)^{1/3}    \left( \frac{n_{\rm ISM}}{1 \rm cm^{-3}} \right)^{-1/3} \left( \frac{\gamma_0}{30} \right)^{-\frac{8}{3}}
\end{equation}

For a wind ($k=2$) this scaling is

\begin{equation}
    t_{\rm p} \sim 10^4 s \left( \frac{E_{\rm iso}}{10^{53} \rm ergs} \right)    \left( \frac{ A_{\rm wind} }{ 5 \times 10^{11} \rm g/cm} \right)^{-1} \left( \frac{\gamma_0}{30} \right)^{-4} .
\end{equation}

These are just estimates; we will compute the coefficients explicitly in this study.  We only note here that the steep dependence on $\gamma_0$ and relatively weak dependence on $E_{\rm iso}$ and $n_{\rm ISM}$ or $A$ puts significant constraints on the inferred value of $\gamma_0$.

As the decay of Lorentz factor with time is more shallow in the stratified model than it is for Blandford-McKee, this results in a shallow plateau in the X-Ray light curve, via

\begin{equation}
    F \propto t^{-\alpha}
\end{equation}

with the decay index $\alpha$ dependent on the stratification $s$, the nonthermal electron spectral index $p$, and the external medium power-law index $k$:

\begin{equation}
    \alpha = \left\{
    \begin{array}{l@{\quad \quad}l}
        \frac{-6 - 6s + 5k/2 + 5sk/2 + (p/2)(24 -7k + sk)}{2(7 + s - 2k)}
        & \nu_m < \nu < \nu_c \\
        \\
        \frac{-4 -4s + k + sk + (p/2)(24 -7k + sk)}{2(7 + s - 2k)} & \nu > \nu_c \\
    \end{array} \right.
    \label{eqn:alphas}
\end{equation}

In this work we will mainly focus on X-Ray observations of the plateau.  For X-Rays we will typically assume we are above the cooling break, $\nu > \nu_c$.  As noted by \cite{2026MNRAS.549ag833S}, The limit $s \rightarrow 1$ recovers the Blandford-McKee scalings, and $s \rightarrow \infty$ recovers a piston-like scaling for a jet coasting at constant Lorentz factor.

\section{Numerical Methods}
\label{sec:num}

We solve this relativistic hydroydnamical problem explicitly using a numerical approach.  We evolve the equations of relativisitic hydrodynamics, in conservation-law form:

\begin{eqnarray}
    \partial_\mu (\rho u^\mu) = 0 \\
    \partial_\mu T^{\mu \nu} = 0 \\
    T^{\mu \nu} = \rho h u^\mu u^\nu + P g^{\mu \nu},
\end{eqnarray}
where $P$ is the pressure, $h = 1 + (\epsilon + P)/\rho$ is the specific enthalpy, and $\epsilon$ is the energy density.  We assume an ultra-relativistic equation of state,

\begin{equation}
    P = \epsilon/3.
\end{equation}

Numerical calculations are carried out using the JET code \citep{2011ApJS..197...15D, 2013ApJ...775...87D}. The JET code solves the equations of relativistic gas dynamics using a moving mesh.  The mesh motion allows for evolution of very high Lorentz factors over many orders of magnitude of expansion.  

At these very early times, the jet is acausal $\gamma \gg 1/\theta_j$, meaning that the jet can be modeled as a spherically symmetric outflow with total energy $E_{\rm iso}$.  This allows us to perform our calculations in 1D, enabling very high resolution.  This is valid for all times before the jet break.  Jet spreading will not be modeled or investigated in this study.

\subsection{Initial Conditions}

\begin{figure}
\epsscale{1.15}
\plotone{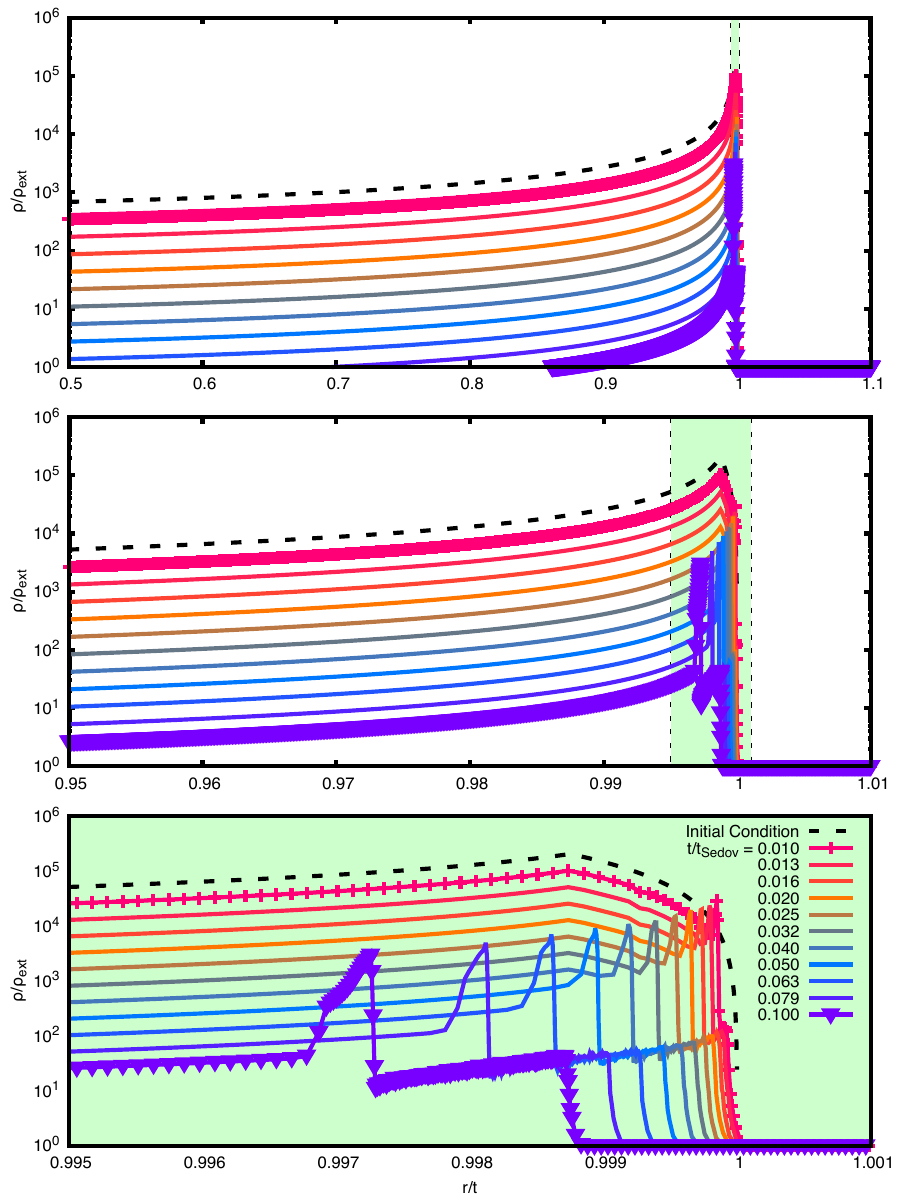}
\caption{ Shock evolution over time for the fiducial ISM case with $\gamma_0 = 20$.  The dashed curve represents the analytical initial condition ($n=2, s=4$), and the colored curves demontstrate the evolution in the shock over an order of magnitude of expansion.  We demonstrate the extremely high resolution necessary using the three panels.  The green shaded region represents the same region of space in all three panels, demonstrating the zoom-in of the shock front and the zone size $\Delta r / r \sim 10^{-5}$ or smaller in the highest-resolved regions.
\label{fig:hires} }
\end{figure}

The density profile of our initial condition is shown in Figure \ref{fig:hires}.  We assume a homologous outflow at some initial time $t$ with velocity given by

\begin{equation}
    v(r) = \left\{
    \begin{array}{l@{\quad \quad}l}
            r/t & r < t \\
            0   & r > t \\
    \end{array} \right.
\end{equation}

The initial conditions assume a stratified ejecta structure with characteristic Lorentz factor $\gamma_0$ (This is equivalent to $\gamma_{\rm min}$ in \cite{2026MNRAS.549ag833S}).  The outer stratification is determined by the index $s$ as defined in Section \ref{sec:anly}.  We also include an index $n$ for the interior density structure, which will largely not effect the afterglow light curve but is included for completeness:

\begin{equation}
\rho_{\rm init}(r) = \left\{ \begin{array}
				{l@{\quad \quad}l}
				\rho_0 (\gamma/\gamma_0)^n & \gamma < \gamma_0 \\  
    			\rho_0 (\gamma/\gamma_0)^{1-s} & \gamma > \gamma_0 \\
                \rho_{ext}(r) & r > c t \\
    			\end{array} \right.
\label{eqn:initrho}
\end{equation}
where

\begin{equation}
\rho_0 = \frac{E_{\rm iso}}{4 \pi t^3 \left[ \frac{1}{n} + \frac{1}{s-1} - \frac{1}{\gamma_0} ( \frac{1}{n-1} + \frac{1}{s} ) \right]}.
\end{equation}
This gives an ejecta with total kinetic energy equal to $E_{\rm iso}$ up to terms of order $1/\gamma_0^2$ \footnote{This normalization gives the correct total energy, including energy moving below $\gamma_0$.  If we wish to make direct comparison with the predictions of \cite{2026MNRAS.549ag833S} we would need to remove the terms involving $n$, as that study took $\gamma_0$ to be the minimum Lorentz factor.  Effectively this would require increasing the energy scale $E_{\rm iso}$ by a factor of $2.4$, adjusting the Sedov time accordingly in order to be consistent with that study.}.

In this study, the external density $\rho_{\rm ext}(r)$ is chosen either to be a uniform ISM density or a wind ($\rho_{\rm ext}(r) = A/r^k$, with either $k=0$ or $k=2$).  The stratification parameter $s$ is taken to be $s=4$ throughout the numerical portion of this work.  Initial pressure is taken to be negligible, $P = 10^{-6}\rho$.  For the ISM calculations ($k=0$), the hydrodynamics are initialized at time $t = 10^{-3} t_{\rm Sedov}$ (lab time).  For wind ($k=2$), the initial time is $t = 10^{-9} t_{\rm Sedov}$.  Our ``code units" assume $E_{\rm iso} = A = c = 1$ so that $t_{\rm Sedov} = 1$, but results are presented assuming the physical values $E_{\rm iso} = 10^{53}$ ergs, $n_{\rm ISM}$ = 1/cm$^3$, $A_{\rm wind} = 5 \times 10^{11}$ g/cm.

\subsection{Resolution}

\begin{figure}
\epsscale{1.15}
\plotone{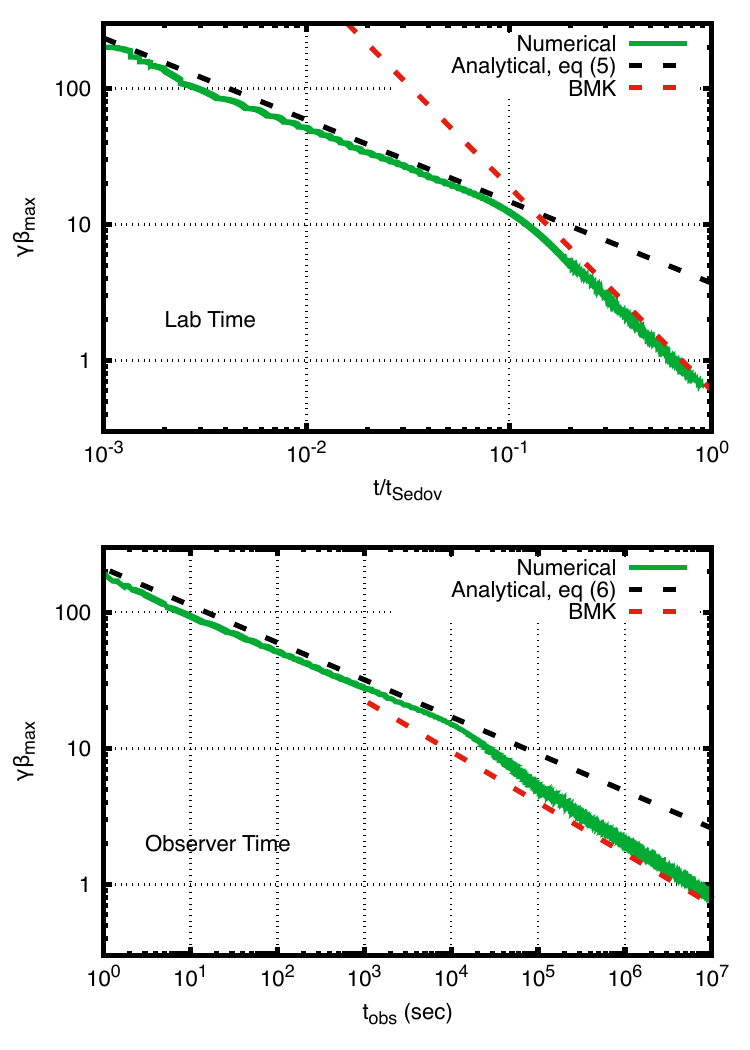}
\caption{ Evolution in the maximum four-velocity on the grid $\gamma \beta_{\rm max}$ over time for the ISM case $k=0$, demonstrating accurate adherence to the analytical scalings (Equation \ref{eqn:anlygam_lab}, \ref{eqn:anlygam}) computed by \cite{2000ApJ...535L..33S} until the reverse shock crosses the ejecta, at which point we recover the standard Blandford-McKee solution \citep{1976PhFl...19.1130B}.
\label{fig:gammab} }
\end{figure}

The \texttt{JET} code has time-variable resolution, with numerical zones dynamically expanding or contracting to react with the flow.  The grid is first initialized with $2 \times 10^4$ computational zones between $r = 0.2t$ and $r = 2 t$.  This is followed by an additional refinement step over any regions where the density gradient is sufficiently steep, increasing the effective resolution by an order of magnitude.  The inner and outer boundaries expand to follow the ejecta, keeping it within $80\%$ of the outer radius.  Finally, active refinement and de-refinement takes place during the course of the calculation, where we only de-refine zones that become compressed by a factor of $20$ smaller than they would be in a purely logarithmic grid.  All of this amounts to a resolution below $\Delta r/r < 10^{-5}$ in the regions requiring the highest resolution (see Figure \ref{fig:hires} for a multiple-zoom-in demonstration of the extremely high resolution being employed in this study).  Figure \ref{fig:gammab} shows that this resolution is sufficient to precisely capture the dynamical evolution of the Lorentz factor with time, accurately evolving Lorentz factors as large as 200.

\subsection{Radiation}
\label{sec:firefly}

We compute synchrotron light curves using the \texttt{Firefly} radiation code \citep{2024ApJ...976..252D, 2025arXiv250415959G}.  \texttt{Firefly} is designed to post-process relativistic hydrodynamics output assuming that electrons become shocked into a standard non-thermal spectrum based on a power-law index $p$, electron energy fraction $\epsilon_e$, and magnetic energy fraction $\epsilon_B$.  \texttt{Firefly} computes a full synchrotron power spectrum assuming a cooling break via a global cooling time $t$, and a characteristic synchrotron break at the frequency $\nu_m$ corresponding to the minimum electron Lorentz factor, but does not account for synchrotron self-absorption in its present version.  \texttt{Firefly} can compute off-axis light curves, spectra, and sky images, but for the present study we will be using it just for the purposes of computing X-ray light curves, assuming on-axis emission.  Synchrotron self-absorption typically dominates only at much lower frequencies, and the absence of this feature in the radiation code does not affect the X-ray light curves. For the radiation, we assume standard typical GRB jet parameters $E_{\rm iso} = 10^{53}$ ergs and $n_{\rm ISM} = 1$ cm$^{-3}$.  For a wind-like external medium, $\rho = A/r^2$, we use $A = 5 \times 10^{11}$ g/cm as our fiducial value.

\section{Results} \label{sec:results}

Our fiducial calculation uses $\gamma_0 = 20$, which will be sufficient to ensure the forward shock is ultra-relativistic as the reverse shock crosses the ejecta.  The maximum value of $\gamma_0$ we investigate is 40, and the necessary resolution scales as $\Delta r/r \sim 1/\gamma_0^2$; if we wished to resolve the solution for $\gamma_0 = 100$, this would require a factor of 6 higher resolution, which would be significantly more expensive.  Nevertheless, our fiducial value $\gamma_0 = 20$ is sufficient to demonstrate the plateau, reverse shock crossing, and transition to blastwave.  Because of self-similarity, the analytical scalings we compute will be extendable to much higher characteristic Lorentz factors if necessary.

Figure \ref{fig:hires} shows the initial condition for $\rho(r,t)$ (Eq. \ref{eqn:initrho}), along with a number of snapshots in the evolution of the shock structure as the reverse shock travels through the stratified ejecta ($k=0,s=4$ case).  This figure demonstrates the extremely high resolution that is necessary to capture the shock evolution; the lower panel is zoomed in on the outer 0.5\% of the shell, which includes the peak at $\gamma = \gamma_0$, at $r/t = 1-1/(2\gamma_0^2) = 0.99875$.

Figure \ref{fig:gammab} presents the evolution in the maximum four velocity $\gamma \beta$ of the outflow, as a function of lab time $t$ (upper panel) and observer time $t_{\rm obs}$ (lower panel).  Analytical scalings (\ref{eqn:anlygam}) first computed by \cite{2000ApJ...535L..33S} are plotted alongside the numerical results, finding precise agreement.  This provides strong evidence that our results are converged.  We also performed separate convergence tests (not shown here) which confirmed that this resolution was necessary to capture the earliest (most relativistic) phases of the evolution.  At a time of $t/t_{\rm Sedov} \approx 0.1$, corresponding to an observer time of $t_{\rm obs}/t_{\rm Sedov} \approx 10^{-4}$, the four velocity $\gamma \beta_{\rm max}$ crosses $\gamma_0$, which means that the reverse shock crosses the bulk of the flow a this stage, resulting in a transition to the standard Blandford-McKee blastwave solution, in which the ejecta mass is considered negligible.

\begin{figure}
\epsscale{1.15}
\plotone{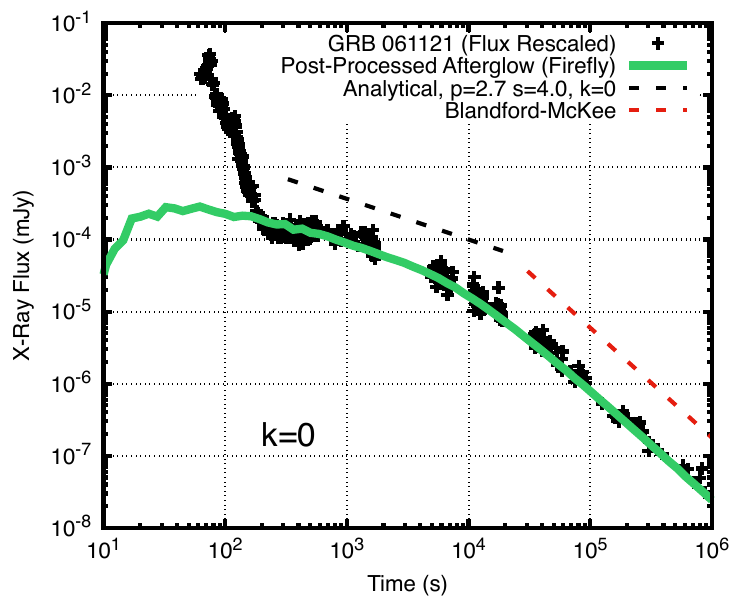}
\caption{ Example afterglow light curve computed from \texttt{Firefly}, compared against the light curve of GRB 061121 as a demonstration of the model fit to an afterglow plateau.  The electron spectral slope $p=2.7$ was chosen to match the slope of the late-time observations.  This afterglow light curve assumed the fiducial model, with a uniform ISM density and $\gamma_0 = 20$.
\label{fig:lc} }
\end{figure}

The checkpoints from the entire hydrodynamic solution were then post-processed using the \texttt{Firefly} code, to produce light curves (Details of our afterglow models are provided in section \ref{sec:firefly}.  An example light curve is shown in Figure \ref{fig:lc}, alongside an example light curve (GRB 061121) with a similar plateau and break time for comparison.  The value of synchrotron power-law $p=2.7$ in our afterglow model was chosen to match the Blandford-McKee phase to the late-time behavior of GRB 061121.  

\begin{figure}
\epsscale{1.15}
\plotone{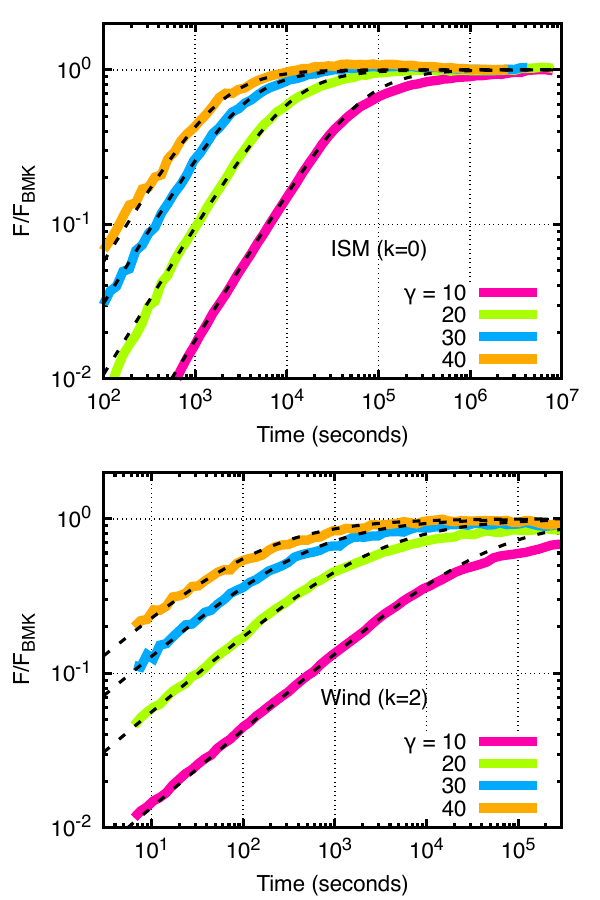}
\caption{ X-Ray afterglow flux curves re-scaled as a fraction of the Blandford-McKee flux, $F/F_{\rm BMK}$, identifying the clear post-plateau break when the flow reaches $\gamma \sim \gamma_0$, after which $F = F_{\rm BMK}$.  The light curves are well-fit by a smoothed broken power-law (Equation \ref{eqn:powerlaw}), where the smoothing parameter $m = 1.5$.
\label{fig:Fbmk} }
\end{figure}

Figure \ref{fig:Fbmk} shows eight re-scaled light curves with different choices of $\gamma_0$ and $k$.  We plot $F/F_{\rm BMK}$ to clearly identify the plaateau break time $t_p$, as a function of $\gamma_0$ and $k$.

We then fitted our afterglow with a broken power-law model:

\begin{equation}
    F_{\rm fit}(t) = \frac{F_p}{\left( (t/t_p)^{(m ~\alpha_s)} + (t/t_p)^{(m ~\alpha_{\rm BMK})} \right)^{1/m}}
    \label{eqn:powerlaw}
\end{equation}

where $\alpha_s$ and $\alpha_{\rm BMK}$ are calculated analytically via equation (\ref{eqn:alphas}) and $F_p$, $t_p$, and the smoothing parameter $m$ are fitted to the light curve via a simple least-squares algorithm.  We use this to find a precise inferred plateau duration for each afterglow.  For $\gamma_0 = 20$, $k=0$, we find $t_p = 1.1 \times 10^4$ seconds.  A full list of plateau durations for all eight of our models is listed in Table \ref{tab:bks}.

\begin{figure}
\epsscale{1.15}
\plotone{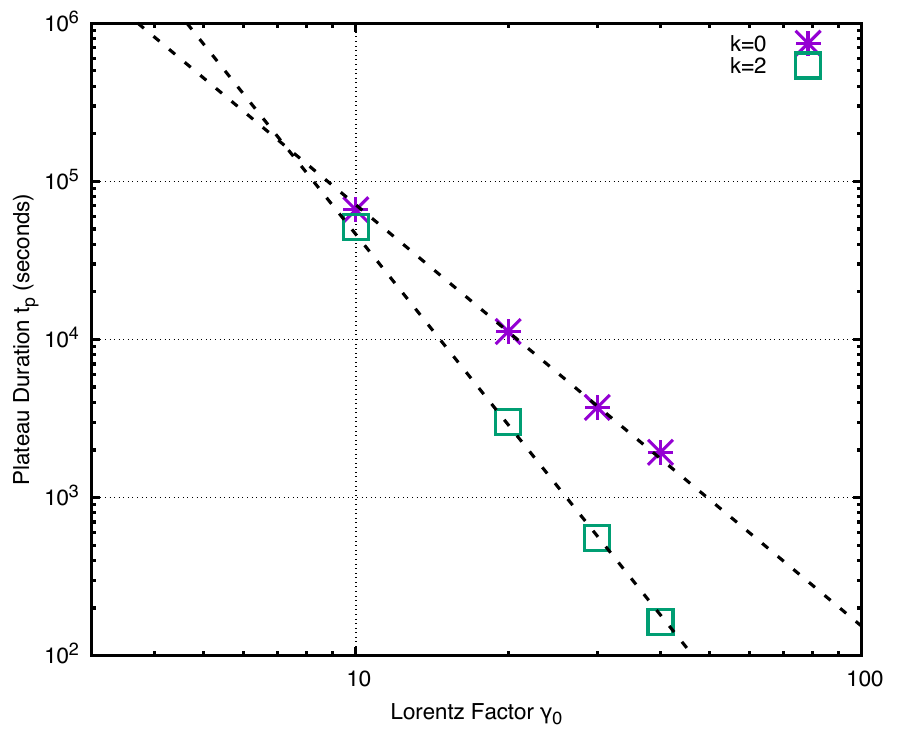}
\caption{ Scalings extrapolated from the break times found in Table \ref{tab:bks}.  As in Table \ref{tab:bks}, we assume $E_{\rm iso} = 10^{53}$ erg for all models. For ISM-like cases ($k=0$), we assume $n_{\rm ISM}$ = 1 cm$^{-3}$.  For wind-like ($k=2$), we assume $A = 5 \times 10^{11}$ g/cm.  Dependence on these parameters can be inferred from analytical scaling arguments.
\label{fig:scaling} }
\end{figure}

The data in Table \ref{tab:bks} are plotted in Figure \ref{fig:scaling}.  By fitting these results we can extrapolate precise coefficients for the plateau break times:

\begin{table}
\begin{center}
    \begin{tabular}{| c c | c | c |}
    \hline
       $\gamma_0$ & $k$ & Plateau Duration $t_p$ & $t_p/t_{\rm Sedov}$\\
    \hline
    ~10~ & ~0~ & $6.6 \times 10^4$ s & $4.89 \times 10^{-4}$\\
    20 & 0 & $1.1 \times 10^4$ s & $8.29 \times 10^{-5}$\\
    30 & 0 & $3.7 \times 10^3$ s & $2.75 \times 10^{-5}$\\
    40 & 0 & $1.9 \times 10^3$ s & $1.43 \times 10^{-5}$\\
    \hline
    10 & 2 & $5.1 \times 10^4$ s & $6.90 \times 10^{-6}$ \\
    20 & 2 & $3.0 \times 10^3$ s & $4.03 \times 10^{-7}$\\
    30 & 2 & $5.5 \times 10^2$ s & $7.44 \times 10^{-8}$ \\
    40 & 2 & $1.6 \times 10^2$ s & $2.19 \times 10^{-8}$ \\
    \hline
    \end{tabular}
    \caption{Measured break times from the light curves generated by eight different \texttt{Firefly} models.  For all plateau durations in this table, we assume $E_{\rm iso} = 10^{53}$ erg. For ISM-like cases ($k=0$), we assume $n_{\rm ISM}$ = 1 cm$^{-3}$ ($t_{\rm Sedov} = 1.4 \times 10^8$ seconds).  For wind-like ($k=2$), we assume $A = 5 \times 10^{11}$ g/cm ($t_{\rm Sedov} = 7.4 \times 10^9$ seconds).
    \label{tab:bks}}
\end{center}
\end{table}

\begin{multline}
    t_{\rm p} = 1.12 \times 10^4 {\rm s} ~(1+z) \\ \times \left( \frac{E_{\rm iso}}{10^{53} \rm ergs} \right)^{1/3}    \left( \frac{n_{\rm ISM}}{1 \rm cm^{-3}} \right)^{-1/3} \left( \frac{\gamma_0}{20} \right)^{-\frac{8}{3}}
\end{multline}

\begin{multline}
    t_{\rm p} = 2.89 \times 10^3 {\rm s} ~(1+z) \\ \times \left( \frac{E_{\rm iso}}{10^{53} \rm ergs} \right)    \left( \frac{ A }{5 \times 10^{11} \rm g/cm } \right)^{-1} \left( \frac{\gamma_0}{20} \right)^{-4}
\end{multline}

Thus, the stratified density structure enables a long plateau, but (in agreement with most other plateau models) we require the characteristic Lorentz factor of our flow to be $\gamma_0 \sim 20$, in the low-to-mid tens.  We note that the ejecta should still have much more relativistic components with $\gamma \gg \gamma_0$ (this is of course an assumed fact of this model) but these highly relativistic components typically only represent a fraction of the total mass and energy.

A more general version of this formula that fits all points (both values of $k$) to within 15\% everywhere is simply

\begin{equation}
    t_p/t_{\rm Sedov} = (1+z)~ 3 \times (2.6 ~ \gamma_0)^{-\frac{8-2k}{3-k}},
\end{equation}
where again

\begin{equation}
    t_{\rm Sedov} = (E_{\rm iso}/A)^{\frac{1}{3-k}}.
\end{equation}

\section{Discussion}

We have performed relativistic hydrodynamics at extremely high resolution in order to capture the evolution of forward and reverse shocks in a stratified ejecta colliding with a circumburst density environment with either a uniform density $\rho =$ constant or wind $\rho = A/r^2$.  We compared the ultra-relativistic shock evolution with established analytical scalings, finding excellent agreement.  We additionally post-processed our hydrodynamic output with the synchrotron radiation code \texttt{Firefly} to produce light curves that are also consistent with analytical predictions.

\subsection{Discrepancies from Previous Work}

Our results are not identical to those of \cite{2026MNRAS.549ag833S}, but may be consistent if a few discrepancies are accounted for.  First, there is a factor of 2.4 in the energy normalization already mentioned previously, due to the fact that our study additionally had a substantial amount of material moving below $\gamma_0$.  Secondly, we predict a fairly broad transition (arguably more than an order of magnitude in time) between the plateau and the BMK phase, which allows for broad interpretation as to when exactly one transitions from one to the other.  \cite{2026MNRAS.549ag833S} computed a specific observer time of reverse shock crossing whereas we fit our afterglow with a broken power-law, which would certainly give different answers.

\subsection{Low Inferred Lorentz Factors}

Our numerical results find plateau durations of $10^4$ seconds only when the outflow's characteristic Lorentz factor $\gamma_0$ is small enough.  Inverting the above expressions we can find formulas for $\gamma_0$ as a function of the break time: 

\begin{multline}
    \gamma_0 = 20.9 \\ \times\left( \frac{t_p / (1+z)}{10^4 \rm s}\right)^{-3/8} \left( \frac{E_{\rm iso}}{10^{53} \rm ergs}\right)^{1/8} \left( \frac{n_{\rm ISM}}{1 \rm cm^{-3}} \right)^{-1/8}.
\end{multline}

For a wind,

\begin{multline}
    \gamma_0 = 14.7 \\ \times \left( \frac{t_p/(1+z)}{10^4 \rm s}\right)^{-1/4} \left( \frac{E_{\rm iso}}{10^{53} \rm ergs}\right)^{1/4} \left( \frac{A_{\rm wind}}{5 \times 10^{11} \rm g/cm} \right)^{-1/4}.
\end{multline} 

Thus to achieve plateaus in the range of $10^3 - 10^5$ seconds, we need Lorentz factors in the range of $\gamma_0 \sim 8.8-50$.  For a wind, the range is narrower, $\gamma_0 \sim 7.4-23$.  Note that $\gamma_0$ is {\em not} the maximum Lorentz factor; in fact, it is the minimum Lorentz factor of the bulk of the ejecta.  Thus, there is a non-zero amount of energy in Lorentz factors $\gamma \gtrsim 100$:

\begin{equation}
    E_{\gamma>100} \approx E_{\rm tot} (\gamma_0/100)^{s-1}.
    \label{eqn:efficiency}
\end{equation}

For $s=4$ and $\gamma_0 = 20$, this gives a fraction of $0.008$ of the ejecta with $\gamma > 100$.  If this component of the ejecta produces gamma rays, this provides an estimate for the prompt efficiency; i.e. the ratio of $E_{\rm iso,\gamma}$ (energy in gamma rays) to $E_{\rm iso}$ (kinetic energy in the jet), which depends on the plateau duration and slope.  Some hints of this have already been seen in observations; \cite{2015ApJ...814....1L} found that cases with significant energy injection also had lower than average prompt efficiencies.  Our specific prediction (\ref{eqn:efficiency}) could be confirmed or ruled out so long as the the parameter $s$ and the kinetic energy $E_{\rm iso}$ are well-constrained.

Unfortunately, the parameter $s$ is typically degenerate with $p$ and $k$, which cannot be constrained from an X-ray light curve alone.  Moreover, the value of the kinetic energy $E_{\rm iso}$ is degenerate with $n_{\rm ISM}$ (or $A$), and other parameters such as $\epsilon_B$ and $\epsilon_e$, which cannot typically be inferred from a single light curve.  The best strategy for confirming this prediction would be a targeted selection of GRB afterglows with well-sampled plateaus in both X-ray and optical, and ideally sufficient spectral information for a multi-band fit.  Constraints from the spectrum would enable a faithful measurement of the parameter $p$, which could be used to identify the BMK phase of the light curve.  The slope of the light curve above the cooling break should be $(3p-2)/4$ (independent of $k$), and the slope below this break depends on $k$, which can be used to constrain the surrounding density profile.  The plateau slope would then be used to determine the value of $s$, and the plateau duration would set constraints on $\gamma_0$.  With sufficient multi-band data (e.g. measurement of a cooling break), one could also constrain some of the nuisance parameters such as $\epsilon_e$ and $\epsilon_B$ that normally make afterglow fitting so highly degenerate.  All of this would be necessary to get a precise measurement of $s$ and $E_{\rm iso}$, which would be necessary to compare this energy with the energy in gamma rays and check the prediction.

If sufficient multi-band afterglow data are available for a single event, and in particular if the values of $E_{\rm iso}$ and $s$ are sufficiently well-constrained, this provides a complete profile for the ejecta which might be sufficient information to develop a reasonable model for the prompt emission; such a model could then be tested against the prompt light curve and spectra.  We intend to investigate this in future work.

Some hints of very low-Lorentz factor ``Dirty Fireballs" with significantly softer spectra have been discovered by the Einstein Probe, with low inferred Lorentz factors based on the plateau duration \citep{2026arXiv260326213D}.  This motivates systematic fitting of afterglows with long plateaus \citep[especially well-sampled afterglows with multi-band data to improve constraints, e.g. ][]{2023A&A...675A.117R, 2024MNRAS.533.4023D}, to search for a potential relationship between plateau duration and peak energy in the prompt emission.  So far, no such correlation has been found.

Alternatively, it is possible that this $\gamma \sim 20$ outflow was produced by the collision of a highly relativistic $\gamma \gtrsim 100$ outflow, with a less-relativistic $\gamma \sim 10$ outflow leading ahead of it (as in \cite{2015ApJ...806..205D}).  In this case, the gamma rays could be produced before the collision, and shock propagation during the collision would naturally generate a modestly relativistic and stratified outflow.  High-resolution numerical calculations of shell collisions could provide insights into natural stratification structures that arise out of these collisions.

\acknowledgments

We thank Anna Ho, Tanmoy Laskar, and Gilad Sadeh for helpful comments and discussions. 

\bibliographystyle{apj} 
\bibliography{jetbib}


\end{document}